\documentclass[dvips,12pt]{article}
\begin{document}


\title{Motion of Spin $1/2$ Massless Particle in
a Curved Spacetime.\\II. Field Lagrangian Approach}

\author{ A.T. Muminov
\\
\footnotesize Ulug-Beg Astronomy Institute, Astronomicheskaya~33,
Tashkent 100052, Uzbekistan\\\footnotesize amuminov2002@yahoo.com
}
\date{}

\maketitle

\begin{abstract}
Earlier we obtained quasi-classical equations of motion of spin
$1/2$ massless particle in a curved spacetime on base of simple
Lagrangian model \cite{al2}. Now we suggest an approach to derive
the equations in framework of field theory. Noether theorem
formulated in terms of Cartan' formalism of orthonormal frames
gives equations for current of spin of the field and tensor of
stress-energy. It is shown that under eikonal approximation the
above mentioned equations can be reduced to equations for
worldline of the particle and equation of spin of the particle
along the worldline. This way conformity between corpuscular
considerations of spin $1/2$  massless particle and approach in
framework of spinor field theory in curved spacetime is
demonstrated.
\end{abstract}

\def \vc {\vec}
\newcommand{\T}{\tilde }
\newcommand{\comment}[1]{}
\newcommand{\lrr}[1]{\left(#1\right)}
\newcommand{\lrc}[1]{\left\{#1\right\}}
\def \om {\omega} 
\def \Om {\Omega} 
\def \tsh {{\cal T}}
\def \wdg {\wedge}
\def \vcx {\dot{\vc x }}
\def \dtx {\dot{x}}
\def \dcn {\mbox{D.C. }}
\newcommand{\lcom}[1]{{\left[#1\right.}}
\newcommand{\rcom}[1]{{\left.#1\right]}}
\newcommand{\lacm}[1]{{\left\{#1\right.}}
\newcommand{\racm}[1]{{\left.#1\right\}}}
\newcommand{\cc}[1]{\bar{#1}}

\newcommand{\ihlf}{\frac{i\hbar}{2}}
\newcommand{\hhlf}{\frac{\hbar}{2}}
\newcommand{\ihqr}{\frac{i\hbar}{4}}
\newcommand{\ihf}{{i\hbar/2}\,}
\newcommand{\hhf}{{\hbar/2}\,}
\newcommand{\ihr}{{i\hbar/4}\,}
\def \qrt {{1/4}\,}
\def \hspin {{1\over2}}
\def \hsp {1/2\,}
\newcommand{\eps}{\varepsilon}
\newcommand{\fracm}[2]
{{\displaystyle#1\over\displaystyle#2\vphantom{{#2}^2}}}
\newcommand{\prt}{\partial\,}
\def \oh {{\tiny 1/2}\,}
\def \as {\!{}^*\!}
\def \Lge {{\cal L}}
\def \vn {{\vec{n}}}
\def \cdt {{\hspace{0.1em}\cdot\hspace{0.1em}}}
\def \fdt {\hspace{0.4em}}
\def \ffdt {\hspace{0.75em}}
\newcommand{\con}[2]{\omega_{#1\cdt}^{\fdt#2}}
\newcommand{\cur}[2]{\Omega_{#1\cdt}^{\fdt#2}}
\newcommand{\Der}[2]{D_{#1}\left(#2\right)}

\newcommand{\dd}[2]{\frac{\prt\,#1}{\prt\,#2}}
\newcommand{\pp}[2]{{\prt\,#1}/{\prt#2}}
\newcommand{\dds}[1]{\frac{d\,#1}{ds}\,}
\newcommand{\Ds}[1]{\frac{D\,#1}{ds}\,}
\def \mtrx #1 #2 #3 #4  {\left(\begin{array}{cc}
#1&#2\\
#3&#4\\
\end{array}\right)}
\def \clmn #1 #2  {\left(\begin{array}{c}
{#1}\\
{#2}\\
\end{array}\right)}
\def \frm #1  {\{#1\}\,{}}
\def \vep {\vc{{}\,\eps}\,}
\def \ps {\Psi}
\def \bps {\bar{\ps}}
\def \gps {{\ps^\dagger}}
\def \dps {D_a{\ps}\,}
\def \dbps {D_a{\bps}\,}
\def \gam {{\hat{\gamma\,{}}\!{}}}
\def \sig {{\hat{\sigma\,{}}\!{}}}
\def \HLC {{\hat{H\,{}}\!{}}}
\def \PR {{\hat{P}}} \def \UN {\hat{1}}
\newcommand{\vel}[1]{{\bps\gam^{#1}\ps}}
\def \elm {electromagnetic}
\section{Introduction}

~~Complete description of motion of a non scalar wave in
gravitational field is given by its covariant field equations.
Under quasi-classical consideration when wave length is
sufficiently less than typical scales of observations, propagation
of the wave is substituted by motion of a particle. Besides, often
it is assumed that such a particle moves along geodesic line.
Frenkel was first who pointed to fact that spin changes trajectory
of motion of particles in external field\cite{frnk}. Motion of
extended spinning particle in curved spacetime was studied by
Papapetrou\cite{papa} and Dixon\cite{Dix}. Similar problem was
studied by Turakulov\cite{zfr} by means of classic Hamiltonian
formalism in approximation of spinning rigid body in tangent
space. In the mentioned works it was shown that equation of motion
differs from geodesic equation due to term which is contraction
the curvature with spin and velocity. A number of attempts to
describe motion of quantum particles with spin on base of
Lagrangian models were made for last eight decades\cite{frszk}.
However, a satisfactory description was not obtained\cite{zr1}.
Nevertheless, studies both Maxwell and Dirac equations point to
the fact that equations of motion might include contraction spin
with the curvature\cite{d16,d17}.

An approach to derive the equations of motion for spin $1$ massive
and massless quantum particles by means of classical Lagrange
formalism are shown in our paper\cite{zr1}. Recently, we extend
area of the studies to spin $1/2$ fields\cite{al,al2} by
constructing classical Lagrangian models describing motion of spin
$1/2$ massive and massless particles. It was shown there that
quasi-classical equations of motion of such particles have form
analogous to Papapetrou equations obtained us earlier for spin $1$
particles.

Since complete description of motion of a non scalar wave in
curved spacetime is given by its covariant field equations it is
evident that a confirmation of the results obtained in framework
of classical Lagrange formalism should be given by consideration
in framework of field theory. Such a confirmation for spin $1$
massless field was achieved in our work\cite{za}, where a
derivation of Papapetrou equations for photon on base of field
variational principles was completed.  Preliminary analysis showed
that straightforward extent of approach\cite{za} developed for
photons to spin $1/2$ fields meets serious mathematical
difficulties. In our opinion, this is caused by significant
difference between nature of fields with noninteger spin and spin
$1$ fields. Goal of present work is to modify a field
approach\cite{za} to derive quasi-classical equations of motion of
particle with spin $1/2$. Currently we confine our considerations
by spin $1/2$ massless field.

Spin $1/2$ field is described by elements of spinor fibre bundle
which consist of local spinor spaces. Determination of the spaces
demands specifying an orthonormal frame in considered domain of
spacetime. We introduce an orthonormal frames especial way which
allows to obtain quasi-classical equations of motion in simplest
form. For this end we specify field of comoving orthonormal frames
such a way that 0-vectors $\vn_-$ are tangent to 0-curves of the
way propagation, and complimentary to them 0-vectors $\vn_+$
($<\vn_-,\vn_+>=1$) specify oscillation of the wave. In turn,
vectors $\vn_\alpha,$ $\alpha=1,2$ describes polarization of the
wave. Procedure of definition the frame is analogous to the one
described in our work\cite{za}. Local spinor spaces are
constructed as follows. Dirac matrices $\frm \gam^a $ referred to
the frame $\frm \nu^a $ are introduced. The matrices generate
local Clifford algebras referred to the frame. Besides, local
Clifford algebra introduced this way specifies two local spinor
spaces referred to the frame. These spaces are two spaces of
spinor representations of the group $SO(1,3)$ which are isomorphic
to each other under Dirac conjugation \cite{ccl,Besse}. Union of
the local spinor spaces constitute spinor fibre bundle in
considered domain of spacetime.
\comment{Since the wave field propagates along some congruence of
0-curves, it is convenient to specify field of comoving
orthonormal frames such a way that 0-vectors $\vn_-$ are tangent
to the 0-curves, and complimentary to them vectors $\vn_+$ specify
oscillation of the wave\cite{za}. In turn, vectors $\vn_\alpha,$
$\alpha=1,2$ describes polarization of the wave. Vector frame
$\frm {\vn_a} ,$ $a=\pm,1,2$ determines dual to it covector frame
$\frm {\nu^a} .$}

Unlike the approach shown in\cite{za}, this time we put
constraints to spinor field after variation the field Lagrangian.
For this we consider partial differential equations of covariant
conservation of current of spin (CS) and stress-energy tensor
(SET) given by Noether theorem formulated in terms of Cartan
theory of orthonormal frames. Then, as before, we specify locally
plane monochromatic (LP) spinor wave field. Consideration of
zeroth order eikonal approximation of Dirac equation under
constraints imposed clarifies structure of sought solution by
extraction spinor fields of definite helicity. Applying the
constraints to conservation equation for current of spin yields
equation for "$-$" component of the current which could be
interpreted as equation for spin of the particle. As it was
expected the equation become identical to equation for spin
obtained under corpuscular consideration\cite{al2}.

Analysis of conservation equation for the SET shows that under
constraints imposed and some additional requirements which has
evident physical meaning the equation is reduced to equation for
generalized momentum (conjugated with coordinates $x^i$) obtained
under corpuscular approach\cite{al2}. This way, conformity between
corpuscular considerations of spin $1/2$  massless particle and
approach in framework of spinor field theory in curved spacetime
has been shown.

\comment{ It should be noted that even the above applied
requirements has obvious physical meaning they may be not
indisputable from point of view of geometry of spacetime.
Nevertheless we daresay that the equation for momentum obtained
under the requirements correctly represents behavior of worldline
of propagation of the wave. Difficulties appeared under derivation
the equation, in our opinion, conditioned by unreality notion of
monochromatic wave. For correct derivation equations of motion of
quantum particle with non integer spin in curved space time in
framework of field theory it is necessary to consider wave packet
characterizing by definite spectrum of energies. However, that is
much more difficult procedure concerning with necessariness of
calculating average momentum, energy and trajectory of the packet.
This is why under first stage of consideration the problem we
prefer to have deal with monochromatic wave.}




\section{Field formalism and Noether invariants}

Spin $1/2$ field is described by elements of spinor fibre bundle
which consist of local spinor spaces. Determination of the spaces
requires specifying an orthonormal frame in considered domain of
spacetime. Besides, condition of orthonormality should be
understood in extended sense. Namely, if $\frm \nu_a $ is such an
orthonormal frame scalar products of the vectors:
$$
<\nu^a,\nu^b>=\eta^{ab},
$$
should constitute an invertible matrix $(\eta^{ab})$ with constant
elements.

In order to describe spinor fields which are arguments of field
Lagrangian we introduce Dirac matrices $\frm \gam^a $ referred to
the frame $\{\nu^a\}$. The matrices are constant in chosen frame
and obey anticommutation rules as follows:
\begin{equation}\label{gam-rules}
\{\gam^a,\gam^b\}=2\eta^{ab}.
\end{equation}
Algebraic span of Dirac matrices yields local sample of Clifford
algebra. Union of the local Clifford algebras constitute fibre
bundle of Clifford algebra in considered domain of spacetime.

Invertible elements $R$ of Clifford algebra such that
$$ R^{-1}=\bar{R},
$$where $\bar{R}$ stands for Dirac conjugated matrix, constitute
$Spin(1,3)$ group \cite{ccl}. There is an endomorphism
$R:SO(1,3)\to Spin(1,3)$ defined by formula:
\begin{equation}\label{trans}
R_L\gam^a R^{-1}_L=L_{b\cdt}^{\fdt a}\gam^b,\quad (L_{b\cdt}^{\fdt
a})\in SO(1,3).
\end{equation}
Elements of local Clifford algebra are operators on two local
spinor spaces referred to the frame. The spaces are local linear
spaces of representation of group $Spin(1,3)$ and $SO(1,3)$.
Elements of the local spaces $\psi\in S$ and $\bps\in \bar{S}$ are
$4\times1$ and $1\times4$ complex matrices accordingly.  This way
element $L$ of Lorentz group $SO(1,3)$ acts on spaces of
representation of the group as follows:
\begin{equation}\label{spstrans} '\ps=R_L\psi,\quad
'\bps=\bps R_L^{-1},\quad \ps\in S,\,\bps\!\in \bar{S}.
\end{equation}
Union of the local spinor spaces constitute spinor fibre bundle in
considered domain of spacetime. Covariant derivatives of the
spinor fields take into account transformations of spinor group:
\begin{eqnarray}\label{der}
D_a\ps=\vn_a(\ps)-\qrt\gamma_{abc}\,\gam^b\gam^c\ps,\\
D_a\bps=\vn_a(\bps)+\qrt\gamma_{abc}\,\bps\gam^b\gam^c,
\end{eqnarray}
where Cartan' rotation 1-forms
$\con{b}{a}=\gamma_{cb\cdt}^{\fdt\fdt a}\nu^c$ specify rotations
of the frame.

\comment{ Image of an infinitesimal rotation $L={\bf1}-\eps\in
SO(1,3)$ is:
\begin{equation}\label{RInf}
R_{1-\eps}=\hat{1}+\delta Q=\hat{1}-1/4\,\eps_{ab}\,\gam^a \gam^b.
\end{equation}
The infinitesimal transformation rotates elements of the rest
frame:
\begin{equation}\label{rot-cvec}
\delta\nu^a=-\eps_{b\cdt}^{\fdt a}\nu^b.
\end{equation}
Accordingly (\ref{spstrans}) the rotation initiates a
transformation of spinors:
\begin{equation}\label{spinRot}
\delta\ps=-1/8\,\eps_{bc}\,[\gam^b,\gam^c]\,\ps,\quad
\delta\bps=1/8\,\eps_{bc}\,\bps\,[\gam^b,\gam^c],
\end{equation}
under which  Dirac matrices rotates as follows:
\begin{eqnarray*}
'\gam^a=R\gam^a R^{-1},\quad \delta\gam^a=[\delta Q,\gam^a];\\
\delta\gam^a=-\eps_{b\cdt}^{\fdt a}\gam^b=
-1/4\eps_{bc}\left[\gam^b\gam^c,\gam^a\right].
\end{eqnarray*}
It is seen that the rotation coincides with rotation of components
of contravariant vector with accuracy up to opposite sign. Thus,
if we take into account both of the transformations Diirac
matrices become invariant \cite{Besse}.}

Field Lagrangian of spin $1/2$ massless field includes spinor
variables $\ps(x^i)$, $\bps(x^i)$ and their derivatives referred
to the orthonormal frame $\{\nu^b\}$:
\begin{equation}\label{4lag}
\Lge=\tsh_{ab}<\nu^b,\nu^a>,\quad \tsh_{ab}=
-\ihlf\left\{\bps\gam_bD_a\ps-D_a\bps\gam_b\ps\right\}.
\end{equation}
It should be noted that Dirac matrices which have invariant form
in given frame also have zero covariant derivative (under taking
into account their external index). Action of the field specified
by the Lagrangian does not depend on choice of the frame:
$${\cal A}=\int L\,\epsilon,\quad
\epsilon={1\over4!}\,{}\,\eps_{abcd}\,{}\,\nu^a\wdg\nu^b\wdg\nu^c\wdg\nu^d.
$$

\comment{In terms of 4-form of the Lagrangian the field equations
have a form as follows:
\begin{equation}\label{feq}
{\prt\Lambda\over\prt\ps}-D_a{\prt\Lambda\over\prt\dps}=0, \quad
{\prt\Lambda\over\prt\bps}-D_a{\prt\Lambda\over\prt\dbps}=0,
\end{equation}
that gives Dirac equations for massless field:
\begin{equation}\label{Dir}
\gam^a\,D_a\ps=0,\quad D_a\bps\,\gam^a=0.
\end{equation}
}

It is known that variational principles is somewhat universal tool
to obtain equations for physical fields. For instance, under
infinitesimal change of metric structure of spacetime described by
variations of orthonormal frame $\frm \nu^a $ we obtain Einstein
equations. Particularly such a variation of the action of spin
$1/2$ field gives stress energy tensor (SET) of the field:
\begin{equation}\label{set}
T_{ab}=-\tsh_{ab}+\tsh_{c\,\cdt}^{\fdt c}\,\eta_{ab},
\end{equation}
Besides SET of the spin $1/2$ field is not symmetric \cite{NN}, so
does $\tsh_{ab}$ which for simplicity we also call SET.

Another type of variations which do not affect to the geometric
structure of spacetime but changes system of observers is
considered by Noether theorem. Due to Noether theorem each class
of infinitesimal local transformations which left action of the
field invariant leads to some differential equation for the field
observables.

\subsection{Noether theorem and Spin current}

Infinitesimal local rotations of field of the orthonormal frames
$\{\nu^b\}$:
\begin{equation}\label{lor-rot}
\delta\nu^a=-\eps^{\fdt a}_{b\cdt}\nu^b,\quad
\eps_{ab}+\eps_{ba}=0,
\end{equation}
leave action of the field invariant, since the one does not depend
on choice of the frames. Thus, owing to Noether theorem,
differential equation:
\begin{equation}\label{eq4spin-cur}
D_c\left[S_{ab\cdt}^{\fdt\fdt c}\right]=T_{\lcom{a}\rcom{b}},
\end{equation}
including covariant divergence of CS\cite{NN}:
\begin{equation}\label{spin}
S^{abc}={i\hbar\over8}\bps\lrc{\gam^c,\gam^\lcom{a}\gam^\rcom{b}}\ps,
\end{equation}
and commutator of SET of the field should be satisfied.

\subsection{Coordinate variation}

Infinitesimal vector field $\vc\eps $ drags coordinate
hyper-surfaces $\{x^i\}$ onto $\{y^i\}$ and elements of
orthonormal covector and vector frames $\{\vc n_a\},$ $\{\nu^b\}$
onto dragged frames $\{'\vc n_a \},$ $\{'\nu^b\}.$  Such a class
of local infinitesimal coordinate transformations leaves action
invariant and lead to differential equation for covariant
divergence of SET as follows:
$$
D_b\left[T^{cb}\right]=\oh R_{ab\cdt\cdt}^{\fdt\fdt
cd}\,S^{ab}_{\cdt\cdt d}.
$$
Proceeding from (\ref{spin}) and anti-commutation rules for Dirac
matrices, we find that CS is double antisymmetric over the
indexes:
\begin{equation}\label{persc}
S^{abc}=S^{bca}=S^{cab}.
\end{equation}
Owing Bianchi identity this reduces RHS of the equation for SET
divergence to zero:
\begin{equation}\label{conset}
D_b\left[T^{cb}\right]=0.
\end{equation}


\section{Canonical frame and structure of spinor wave}

We confine our consideration by spinor wave fields whose
propagation in some 4-tube in a curved spacetime can be described
as motion of particle with spin. Namely, the wave propagates along
some congruence of 0-curves inside 4-tube representing 0-worldline
of the particle. The wave in 4-tube is determined by field of
orthonormal frames (associated with the congruence) and components
of spinors referred to the frame. Since the wave admits
corpuscular description we assume that in first order eikonal
approximation interaction of the spinor field with gravitation
expresses shape of the 0-curves (to be exact dynamic of the field
of orthonormal frames), whereas components of spinors satisfy
simple equations similar to analogous equations in flat spacetime.
It should be noted that such situation is provided by choice of
suitable orthonormal frame.  Procedure of choice of such a frame
for electromagnetic wave propagating in curved spacetime was
described in our earlier work \cite{za}. However it is relevant to
explain analogous procedure for the case in question more detail.
First of all we consider plane wave of massless spinor field in
Minkowski spacetime.

\subsection{Plane wave in Minkowski spacetime}

Flat space time admits cartesian coordinates $\{x^i\}$ and by
suitable transformation of the coordinates we can put that
direction of $x^3$ coordinate axis is coincide with direction of
propagation of the wave, whilst $x^0=ct$ is time coordinate. This
way we obtain an orthonormal vector frame $\{\vn_i\}=\{\prt/\prt
x^i \}$. In turn, dual to the frame covector frame $\nu^a$
specifies spinor fiber bundle in the space by map $T^*\to CL(1,3)$
which specified by its images on elements of frame of cotangent
space: $\nu^a\to\gam^a$. The images are Dirac matrices. In chiral
representation they take matrix form as follows \cite{ccl}:
\begin{eqnarray}\label{chiralREP}
\gam^0=\mtrx 0 1 1 0 ,\quad\gam^\mu=\mtrx 0 {\sig^\mu} {-\sig^\mu}
0 ,\, \mu=1,2,3;\\\nonumber \sig^1=\mtrx 0 1 1 0
,\quad\sig^2=\mtrx 0 -i i 0 ,\quad \sig^3=\mtrx 1 0 0 -1 ;
\end{eqnarray}
where $\sig^\mu$ are Pauli matrices. Since we consider spinor
plane wave $\ps$ propagating in $\vn^3$ direction, massless Dirac
equation for spinor becomes:
$$0=\gam^a\vn_a\circ\ps=(\gam^0\vn_0+\gam^3\vn_3)\circ\ps.
$$
The equation gets simplified if we formulate the one in terms of
null vectors:
\begin{eqnarray*}
\nu^+=\nu^0-\nu^3,\quad\nu^-=1/2(\nu^0+\nu^3);\\
\vn_+=1/2(\vn_0-\vn_3),\quad\vn_-=\vn_0+\vn_3.
\end{eqnarray*}
Actually, frames $\{\vn_\pm,\vn_\alpha\}$ and
$\{\nu^\pm,\nu^\alpha\},$ $\alpha=1,2$ are dual to each other and
orthonormal in sense:
\begin{equation}\label{orthFrame}
<\nu^\pm,\nu^\pm>=<\nu^\pm,\nu^\alpha>=<\nu^1,\nu^2>=0,
\quad<\nu^+,\nu^->=-<\nu^\alpha,\nu^\alpha>=1.
\end{equation}
So, the map between Grassman and Clifford algebras may be
equivalently specified by images of generalized Dirac matrixes
referred to frame $\{\nu^\pm,\nu^\alpha\}$:
$$\nu^+\to\gam^+=\gam^0-\gam^3,\quad\nu^-\to\gam^-=1/2(\gam^0+\gam^3);
$$whilst $\gam^\alpha$ remain unchanged. Now
Dirac equation for plane spinor wave is written as:
$$(\gam^+\vn_++\gam^-\vn_-)\circ\ps=0. $$
Since dependence of the spinor plane wave on $x^3,t$ variables is
given by their combination $ct-x^3$, i.e.
\begin{equation}\label{anzatz}
\ps=\exp(i\varphi)\ps_c,\quad \varphi=\omega(t-x^3),\quad
\ps_c=const
\end{equation}
only derivative along vector $\vn_+$ contributes to the equation
for plane spinor wave which due to (\ref{anzatz}) is reduced to
simple matrix equation:
$$\gam^+\vn_+\circ\ps=i\omega\gam^+\ps=0.
$$
This equation gives us solution of massless Dirac equation in form
of plane spinor wave:
\begin{eqnarray*}
\ps_c=\clmn {\chi} {\phi} ,\quad
(1-\sig^3)\phi=(1+\sig^3)\chi=0;\\
\phi=const\clmn 1 0 ,\quad \chi=const\clmn 0 1 .
\end{eqnarray*}
Besides solutions $\clmn 0 {\phi} $ and $\clmn {\chi} 0 $
describes waves with helicity $\pm1$ accordingly.

\subsection{Spinor fields and Locally plane wave in curved spacetime}

Spin $1/2$ field is described by elements of spinor fibre bundle
which consist of local spinor spaces. Determination of the spaces
demands specifying an orthonormal frame in considered domain of
spacetime. Since the wave field propagates along some congruence
of 0-curves, it is convenient to specify field of comoving
orthonormal (in sense (\ref{orthFrame})) frames such a way that
0-vectors $\vn_-$ are tangent to the 0-curves, and complimentary
to them vectors $\vn_+$ specify oscillation of the wave:
\begin{equation}\label{freq}\ps,_{+}=i\omega\ps,
\end{equation}
where frequency of the oscillations $\omega$ characterizes value
of energy $E=\hbar\omega$ in the frame \cite{za}. In turn, vectors
$\vn_\alpha,$ $\alpha=1,2$ describes polarization of the wave.
Vector frame $\frm {\vn_a} ,$ $a=\pm,1,2$ determines dual to it
covector frame $\frm {\nu^a} .$

It should be noted that pair of vectors $\vn_\pm$ is defined with
accuracy up to arbitrary Lorentzian rotations $SO(1,1)$. Rotating
vectors $\vn_\pm$ such a way that value of energy $E$ in the frame
becomes a constant, we define canonical frame so that vector
$\vn_-$ tangent to the 0-curves becomes an analog of canonical
velocity. The equations of motion in quasi-classical approximation
take their simplest form in such a canonical frame.

Since interaction the wave with gravitation is expressed by shape
of the 0-curves (i.e. by rotations of the canonical frame along
$\vn_-$) we expect that components of spinor have no changes in
spatial directions inside the 4-tube:
\begin{equation}\label{constr}\ps,_\alpha=0.
\end{equation}

\subsection{Eikonal approximation}

Essence of the approximation is assumption that the derivative
$\vn_+\ps=i\omega\ps$ is a dominating derivative. Accordingly
(\ref{constr}) this means:
\begin{equation}\label{eikonal}
|\gamma_{abc}|\ll\omega,\quad
|\ps,_-|\ll\omega|\ps|.
\end{equation}
Substitution this requirements to Dirac equation gives:
$$0=\gam^a\,D_a\ps=i\omega\gam^+\ps\,+o(\omega\ps)=0.
$$
Thus in zeroth order of eikonal approximation we obtain:
\begin{equation}\label{gamPlus}
\gam^+\ps=0,\quad \bps\gam^+=0.
\end{equation}
This condition clarifies structure of LP spinor wave.

\subsection{Structure of the spinor field}

Condition (\ref{gamPlus}) allows to  recognize structure of
components of locally plane spinor wave in the frame. Actually,
let us consider matrix
\begin{equation}\label{gam5}
\gam^5=-i/2\,\gam^1\gam^2[\gam^+,\gam^-]=\mtrx 1 0 0 -1  .
\end{equation}\def \gfive {\mtrx 1 0 0 -1  }
If we put spinor $\ps$, satisfying (\ref{gamPlus}) on the right
side of matrix $\gam^5$, due to (\ref{gam5}) we obtain:
\begin{equation}\label{redGam5}
\gam^5\ps=-i\gam^1\gam^2\ps=-\HLC\ps,\quad \HLC=\mtrx \sig^3 0 0
\sig^3 ,\quad\HLC^2=\UN;
\end{equation} where $\HLC$ is helicity operator. This shows that
locally plane spinor waves may be eigenfunctions of helicity
operator according to eigenvalues $\pm1$. Owing to definition,
only non zero component of CS $S^{12-}$ (with accuracy up to
cyclic permutation of the indexes) can be expressed as averaged
value of operator of helicity $\HLC$ (multiplied to $\hhf$) in
state described by spinor $\ps$:
\begin{eqnarray*}
S^{12-}=\ihf\,\bps\gam^-\gam^1\gam^2\ps=
\ihr\,\gps\gam^+\gam^-\gam^1\gam^2\ps=\ihf\,\gps\gam^1\gam^2\ps=
\hhf\,\gps\HLC\ps.
\end{eqnarray*}
As usual we introduce projectors:
\begin{eqnarray*}\PR_\pm=\UN\pm\HLC\to\UN\mp\gam^5\\
\Rightarrow \PR^2_\pm=\UN\pm2\HLC+\HLC^2=2\PR_\pm,\quad
\PR^+\PR^-=\PR^-\PR^+=\UN-\HLC^2\equiv0,\\
\HLC\PR_\pm=\HLC\mp\UN=\pm\PR_\pm,\quad 1/2(\PR_++\PR_-)=\UN.
\end{eqnarray*}
Two last equations shows that arbitrary spinor $\ps$ can be
presented as sum of spinors $\PR_\pm\ps$ of definite helicity
$\pm$. Now let $\ps$ is spinor representing LP spinor wave, thus
(\ref{gamPlus}) and Dirac equation is satisfied. Hence, we obtain:
\begin{eqnarray*}
\gam^aD_a\PR_\pm\ps=\gam^aD_a(\UN\mp\gam^5)\ps=\gam^a(\UN\mp\gam^5)D_a\ps=\\
=(\UN\pm\gam^5)\gam^aD_a\ps=\PR_\mp\gam^aD_a\ps\equiv0,
\end{eqnarray*}
where we used (\ref{redGam5}) and fact that matrix $\gam^5$
anticommutates with each of Dirac matrices. This way it is shown
that each of component $\ps_\pm=1/2\PR_\pm\ps$ (of helicity $\pm1$
) of spinor $\ps=\ps_++\ps_-$ representing LP spinor wave also
satisfies Dirac equation. Since LP spinor wave admits description
in terms of corpuscular theory which considers spin $1/2$ massless
particle of definite helicity, hereafter we will consider only LP
spinor waves $\ps=\ps_\pm$ of definite helicity:
\begin{equation}\label{HLC}
\HLC\ps=H\ps,\quad H=\pm1.
\end{equation}
Thus, accordingly to (\ref{redGam5}) we can recognize structure of
spinor wave in given representation:
\begin{eqnarray}\label{strucSP}
\ps_+=\clmn 0 {\phi} ,\quad \ps_-=\clmn {\chi} 0 ,\quad
\sig^3\phi=\phi,\\\sig^3\chi=-\chi,\quad\phi=F\clmn 1 0 ,\quad
\chi=F\clmn 0 1 ;\nonumber\\
\ps=F\Psi_c,\quad \Psi_c=\ps_\pm/F;\nonumber
\end{eqnarray} where $F$ some scalar.
Besides, accordingly to our assumption that $\vn_+\ps=i\omega\ps$
is dominating derivative we put:
\begin{equation}\label{F}
F=exp(i\varphi)\,f,\quad \varphi_{,+}=\omega=const,\quad
f_{,+}=0,\quad Im(f)=0.
\end{equation}
Moreover, accordingly (\ref{constr}) we assume:
\begin{equation}\label{f_al}
f_{,\alpha}=\varphi_{,\alpha}=0.
\end{equation}


Structure of spinor allows to find out view of some observables:
\begin{equation}\label{velocity}
\vel{\alpha}=0,\quad \vel{-}=\gps\ps=|f|^2,\quad \vel{+}=0
\Rightarrow \bps\gam^a\ps=\delta^a_-\,|f|^2.
\end{equation}
Let's apply this property to calculation derivative of "-"
component of the "velocity" \cite{mss} of spinor field along
vector $\vn_-$:
\begin{eqnarray*}
\vn_-(\vel{-})=\vn_a(\vel{a})-\vn_\alpha(\vel{\alpha})-\vn_+(\vel{+})=\\
=D_a\bps\,\gam^a\ps+\bps\gam^a\,D_a\ps-\gam_{ac\cdt}^{\fdt\fdt
a}\,\vel{c}=-Div(\vn_-)\vel{-},\quad\mbox{i.e.:}
\end{eqnarray*}
\begin{equation}\label{coneq}
\vn_-|f|^2+Div(\vn_-)\,|f|^2=0.
\end{equation}

\subsection{Structure of stress-energy tensor}

Structure of spinor wave  allows to clarify structure of the SET.
Actually, due to (\ref{strucSP},\ref{F},\ref{f_al}) and
(\ref{tsh_ab}) after simple calculations we find out:
\begin{eqnarray}\nonumber
\tsh_{ab}=E\,\eta_{a-}\,(\bps\gam_b\ps)+1/2\,\gamma_{ade}\,S^{de}_{\cdt\cdt
b}
=Ef\,\eta_{a-}\eta_{b-}+1/2\,\gamma_{ade}\,S^{de}_{\cdt\cdt b},\\
\label{tsh_ab}\mbox{i.e.}\quad
 \tsh_{a\beta}=\gamma_{a-\delta}\,S^\delta_{\cdt\beta},\quad
\tsh_{a-}\equiv0,\quad
 \tsh_{a+}=Ef\,\eta_{a-}+1/2\gamma_{acd}\,S^{cd}.
\end{eqnarray}


\section{Equations of motion}

In this section we derive equations of motion describing
propagation of spin $1/2$ massless wave field in corpuscular terms
from differential equations for Noether invariants.

\subsection{Equation for spin}

Now let's consider nonzero components $S^{abc}$ of CS. Due to
definition and (\ref{persc}) indexes $a,b,c$ should differ each
other. Owing to (\ref{gamPlus}), it is seen that no one of them
may be equal to "+". So only possibility is:
\begin{equation}\label{spinPro}
S^{abc}\neq0\,\Rightarrow\,\{a,b,c\}=\{1,2,-\}.
\end{equation}
Since the wave is propagating in $\vn_-$ direction it is essential
to define spin of the particle describing propagation of the wave
as follows:
\begin{equation}\label{spinDef}
S^{ab}=S^{ab-}.
\end{equation}
It is seen that only nonzero components of the spin are
$S^{12}=-S^{21}.$

Starting from conservation law for current of spin
(\ref{eq4spin-cur}) we derive equation for nonzero components of
the spin:
\begin{eqnarray*}
T^{[\delta\eps]}=-\tsh^{[\delta\eps]}=D_a(S^{\delta\eps
a})=D_a(S^{\delta\eps})^a+\gam_{ac\cdt}^{\fdt\fdt a}S^{\delta\eps
c}=\\
=D_-(S^{\delta\eps})+D_+(S^{\delta\eps})^
++D_\alpha(S^{\delta\eps})^\alpha+Div(\vn_-)\,S^{\delta\eps}=\\
=D_-(S^{\delta\eps})+Div(\vn_-)\,S^{\delta\eps} +
\gamma_{\alpha-\cdt}^{\fdt\fdt\lcom{\delta}}\,{S\vphantom{S'}}^{\rcom{\eps}\alpha},
\end{eqnarray*}
where indexes which stay outside brackets at symbol of covariant
derivation $D$ indicate that under calculation the derivative they
should not be taken into account. Such a manner we obtain:
$$D_-(S^{\delta\eps})+Div(\vn_-)\,S^{\delta\eps}=
-\gamma_{\alpha-\cdt}^{\fdt\fdt\lcom{\delta}}\,{S\vphantom{S'}}^{\rcom{\eps}\alpha}
-1/2\,\gamma^\lcom{\delta}_{\cdt de}\,
{S\vphantom{S'}}^{de\rcom{\eps}}.
$$ Let $\delta=1,$ $\eps=2.$ Noting that all nonzero elements of
$S^{abc}$ are elements with indexes $\{a,b,c\}=\{1,2,-\}$ we see
that RHS of the above equation is zero:
$$-\gamma_{1-\cdt}^{\fdt\fdt1}\,{S\vphantom{S'}}^{21}
+\gamma_{2-\cdt}^{\fdt\fdt2}\,{S\vphantom{S'}}^{12}
-\gamma^1_{\cdt -\,1}\,{S\vphantom{S'}}^{12}
 +\gamma^2_{\cdt -\,2}\,{S\vphantom{S'}}^{21}=0.
$$
This gives us:
\begin{equation}\label{eq4spin}
D_-(S^{\delta\eps})+Div(\vn_-)\,S^{\delta\eps}=0.
\end{equation}
It is convenient to introduce normed spin:
\begin{equation}\label{normSpin}
\sigma^{ab}=S^{ab}/(\bps\gam^-\ps),
\end{equation}
which become an analog of spin $1/2$ particle. Due to
(\ref{coneq}) equation for normed spin simplifies:
\begin{equation}\label{eq4NS}
D_-(\sigma^{\delta\eps})=0.
\end{equation}
The equation coincides with analogous equation for spin $1/2$
massless particle obtained us earlier under consideration Lagrange
model of the particle \cite{al2}.

\subsection{Equation for generalized momentum}

Since SET of quantum wave field has components which are analogs
of momentum of classic particle equation of continuity for SET is
expected to be reduced to equation for momentum of classic
particle under some conditions.  From the other hand, if we obtain
exact solution of Dirac equation and substitute it into the
equation for SET continuity we obtain trivial identity $0=0$. But
essence of our approach is to avoid solving field equation and to
extract necessary information about worldline of propagation of
the wave from the equations for observables (SC and SET). We put
suitable requirements to the field representing motion of the
particle. That allowed to clarify structure of the field.  When we
substitute such a field into the equation for SET this time we
expect to obtain an equation for components of SET which can be
interpreted as equation for momentum of the particle. The equation
should coincide with obtained us earlier equation for generalized
momentum of spin $1/2$ massless particle derived in framework of
classic Lagrange formalism \cite{al2}. Due to the above mentioned
arguments let us analyze continuity equation for SET:
\begin{eqnarray}\nonumber
0=D_b(\tsh_{a\cdt}^{\fdt b})-D_b(\delta_a^b\,\tsh_{c\cdt}^{\fdt
c})=D_b(\tsh_{a\cdt}^{\fdt b})-D_a(\tsh_{c\cdt}^{\fdt c})=\\
\nonumber
=D_b(\tsh_a)^b_\cdt+\gamma_{bc\cdt}^{\fdt\fdt
b}\,\tsh_{a\cdt}^{\fdt c}
-D_a(\tsh_c)^c_\cdt-\gamma_{ab\cdt}^{\fdt\fdt\,
c}\,\tsh_{c\cdt}^{\fdt b}=\\
\nonumber
 =D_-(\tsh_a)^-_\cdt+\gamma_{b-\cdt}^{\fdt\fdt
b}\,\tsh_{a\cdt}^{\fdt-}
-D_a(\tsh_-)^-_\cdt+\Delta=\\
\label{dyneq}=
D_-(\tsh_a)^-_\cdt+Div(\vn_-)\,\tsh_{a\cdt}^--D_a(\tsh_-)^-_\cdt+\Delta;\\
\nonumber
 \Delta=D_\beta(\tsh_a)^\beta_\cdt+\gamma_{b\delta\cdt}^{\fdt\fdt
b}\,\tsh_{a\cdt}^{\fdt \delta}
-D_a(\tsh_\delta)^\delta_\cdt-\gamma_{ab\cdt}^{\fdt\fdt\,
c}\,\tsh_{c\cdt}^{\fdt b}.
\end{eqnarray}
Here we write components $\tsh_{a\cdt}^{\fdt-}$ of SET which can
be interpreted as components of momentum of classic particle
separately. To be exact let us define analog of generalized
momentum of spinor wave field as:
\begin{equation}\label{Pa}P_a=\tsh_{a\cdt}^-/|f|^2=
E\,\eta_{a-}+1/2\,\gamma_{ade}\,\sigma^{de}.
\end{equation}
It is to note that such an analog of the momentum has the same
structure as generalized momentum derived in our later work
\cite{al2} under consideration Lagrange model of spin $1/2$
massless particle. Thus due to (\ref{coneq})
$$D_-(\tsh_a)^-_\cdt+Div(\vn_-)\,\tsh_{a\cdt}^-=|f^2|\,D_-(P_a)
$$
After this equation (\ref{dyneq}) can be rewritten as:
\begin{eqnarray}\label{dyneq2}
-|f|^2\left\{D_-(P_a)-D_a(P_-)\right\}=2|f|_{,a}P_-+\\\nonumber
+\left[D_\beta(\tsh_a)^\beta_\cdt+\gamma_{b\delta\cdt}^{\fdt\fdt
b}\,\tsh_{a\cdt}^{\fdt \delta} -D_a(\tsh_\delta)^\delta_\cdt
-\gamma_{a-\cdt}^{\fdt\fdt\, c}\,\tsh_{c\cdt}^{\fdt-}
-\gamma_{a\beta\cdt}^{\fdt\fdt\,
c}\,\tsh_{c\cdt}^{\fdt\beta}\right].
\end{eqnarray}
It can be demonstrated that LHS of the equation coincides with LHS
of equation for generalized momentum \cite{al2}. Besides, under
$a=-$ the LHS turns to equality. If $a\neq-,$ due to
(\ref{f_al},\ref{F}) $D_a|f|^2=0$ and all the rest terms of RHS of
the above equation can be neglected under some additional
requirements. Actually, if we consider locally plane wave
propagated along some congruency of 0-worldlines it is essentially
to assume that there is some central zone in area of the wave
propagating where infinitesimally neighboring worldlines are
locally parallel to each other. So that field of vectors $\vn_-$
has vanishing rotations in directions $\vn_a,$ $a\neq-$, i.e.
rotation coefficients $\gamma_{a-\cdt}^{\fdt\ffdt b}$ are
negligible. If we accept such requirements, accordingly
(\ref{tsh_ab}), elements $\tsh_{c\beta}$ of SET also become
negligible and RHS of (\ref{dyneq2}) certainly vanishes.

It should be noted that even the above applied requirements has
obvious physical meaning they may be not indisputable from point
of view of geometry of spacetime. Nevertheless we daresay that
equation
\begin{equation}\label{dyneq3}
D_-(P_a)-D_a(P_-)=0,
\end{equation}
obtained under the requirements represents correctly a behavior of
worldline of propagation of the wave. Difficulties appeared under
derivation the equation, in our opinion, conditioned by unreality
notion of monochromatic wave. For quite correct derivation the
equations of motion of quantum particle with non integer spin in
curved space time in framework of field theory it is necessary to
consider wave packet characterizing by definite spectrum of
energies. However, that is much more difficult procedure requires
calculating average momentum, energy and trajectory of the packet.
This is why under first stage of consideration the problem we
prefer to have deal with monochromatic wave.

Anyhow, more or less correct we obtain equation (\ref{dyneq3}) for
analog of generalized momentum (\ref{Pa}). Let us rewrite its
second term in explicit form:
$$
D_a(P_-)=1/2\,\left(\gamma_{-cd}\,\sigma^{cd}\right)_{,a}
-\gamma_{a-\cdt}^{\fdt\fdt\,b}\,P_b=
1/2\,\gamma_{-cd,a}\,\sigma^{cd}-\gamma_{a-\cdt}^{\fdt\fdt\,b}\,P_b,
$$
where we take into account the fact that coefficients
${\sigma^{cd}}$ are constants. After this, finally, equation
(\ref{dyneq3}) takes form as follows
\begin{equation}\label{eq4tsh}
D_-(P_a)+\gamma_{a-\cdt}^{\fdt\fdt\,b}\,P_b
-1/2\,\gamma_{-cd,a}\,\sigma^{cd}=0.
\end{equation}
It is seen that form of the equation coincides with equation for
generalized momentum obtained us earlier in work \cite{al2} under
consideration Lagrangian model for spin $1/2$ massless particle.
Thus, exactly the same manner, this equation can be reduced to
Papapetrou equation for trajectory of the particle derived in our
paper \cite{al2}:
\begin{equation}\label{papa}
E\cdot {D\vphantom{,}}_{\!-}(\vn_-)=\hsp R^{\fdt\fdt
a}_{\delta\eps\cdt -}\,\sigma^{\delta\eps}\,\vn_a,
\end{equation}
where
$$\cur{c}{d}=d\con{c}{d}+\con{e}{d}\wdg\con{c}{e}=1/2R_{c\cdt
ab}^{\fdt d}\,\nu^a\wdg\nu^b
$$ is 2-form of curvature.



\section*{Acknowledgment}
The author thanks to  Z. Ya. Turakulov who motivated the author to
consider this problem and whose valuable remarks provided
significant improve of present article.
%

\end{document}